\documentclass[aps,pra,showpacs,groupedaddress,twocolumn]{revtex4}
\usepackage{eurosym}
\usepackage{amsfonts}
\usepackage{graphicx}
\usepackage{latexsym}
\usepackage{makeidx}
\usepackage{amsmath}
\usepackage{amssymb}

\begin{document}

\title{Dynamical Casimir Effect in cavity with $N$-level detector or $N-1$
two-level atoms}
\author{A. V. Dodonov and V. V. Dodonov}
\affiliation{Instituto de F\'{\i}sica, Universidade de Bras\'{\i}lia, PO Box 04455,
70910-900, Bras\'{\i}lia, Distrito Federal, Brazil}

\begin{abstract}
We study the photon generation from vacuum via the Dynamical Casimir Effect
in a cavity containing a $N$-level detector in equally-spaced resonant
ladder configuration or $N-1$ identical resonant two-level atoms.
If the modulation frequency equals exactly twice the unperturbed cavity
frequency, the photon growth goes on steadily for odd $N$, while for even $N$
at most $N-2$ photons can be generated. This finding is corroborated by
numerical calculations. In the limit $N\to\infty$ we obtain the harmonic
oscillator model of the detector, which admits analytical results.
\end{abstract}

\maketitle

A possibility of creating quanta of the electromagnetic field from the
initial vacuum state in cavities with moving boundaries, called nowadays
as the Dynamical Casimir Effect (DCE), was a subject of numerous theoretical
studies for a long time: see, e.g., the most recent reviews
\cite{revDCE,revDal,revRMP}. It was shown \cite{DK96} that one might expect a significant
rate of photons generation inside ideal cavities with resonantly oscillating
boundaries.
The simplest model describing this effect takes into account
 a single resonant cavity mode whose frequency is
rapidly modulated according to the harmonical law $\omega _{t}=\omega
_{0}[1+\varepsilon \sin (\eta t)]$ with a small modulation depth, $%
|\varepsilon |\ll 1$. We shall use
dimensionless variables, setting $\hbar =\omega _{0}=1$. Then the
Hamiltonian for the resonance mode has the form
\cite{Law94}
\begin{equation}
H_{c}=\omega _{t}n-i\chi _{t}(a^{2}-a^{\dagger 2}),
\quad
\chi _{t}=(4\omega _{t})^{-1}d\omega_{t}/dt,
  \label{Hc}
\end{equation}%
where $a$ and $a^{\dagger }$ are the cavity annihilation and creation
operators, and $n\equiv a^{\dagger }a$ is the photon number operator.
 It is well known that the number of photons created from the
initial vacuum state is maximal if the modulation frequency is
exactly twice the unperturbed mode frequency, i.e., $\eta =2$.
The mean number of photons $\langle n\rangle$ and
the Mandel factor $Q=[\langle (\Delta n)^{2}\rangle -\langle n\rangle
]/\langle n\rangle $  increase with time in this ideal case as
(hereafter we use the subscript $0$ for the quantities related to the empty cavity)
\begin{equation}
\langle n_{0}(t)\rangle =\sinh ^{2}(\varepsilon t/2),
\quad Q_{0}(t)= %\cosh \varepsilon t=
1+2\langle n_{0}(t)\rangle .
  \label{n0}
\end{equation}%
The field mode goes to the squeezed vacuum state
with the following variances of the field quadrature operators
$x=(a+a^{\dagger })/\sqrt{2}$ and $p=(a-a^{\dagger })/(\sqrt{2}i)$
(the average values of these operators are zero)
\begin{equation}
(\Delta p_{0})^{2}=\frac{1}{2}e^{-\varepsilon t},\qquad (\Delta x_{0})^{2}=%
\frac{1}{2}e^{\varepsilon t}.  \label{p0}
\end{equation}%

But simple formulas (\ref{n0}) and (\ref{p0}) hold for the ideal empty cavity only.
To registrate the emerging photons one has to couple the field mode to
some detector. And here the problem of the back action of the detector on the field
arises, because in many realistic cases the coupling between the field and detector
can be much stronger than that between the field and vibrating cavity walls.
This was noticed  in \cite{pla}, where it was shown that for the
simplest model of detector as a two-level ``atom'', no photons can be created at all
for the modulation frequency $\eta=2$, if the field atom coupling constant $g$ is much
bigger than the frequency modulation amplitude $\varepsilon$.
A more detailed investigation of this problem was given recently in \cite{1,2},
where different resonant regimes were found and analyzed for different ratios
$g/\varepsilon$ (both small and big).
On the other hand, in papers \cite{4,5} we have discovered that when the
inter-level transitions are resonant with the cavity bare frequency and the
modulation frequency equals twice the cavity unperturbed frequency,
 many photons can be
generated for the $3$-level atom in ladder configuration or a chain of two $%
2$-level atoms, even if $g\gg \varepsilon$. Therefore, a question naturally arises: how many photons can
be generated if one uses the $N$-level atom in the equidistant ladder configuration
%(or a chain of $N-1$ identical $2$-level atoms),
and the inter-level transitions are resonant with the cavity mode modulated at
exactly twice the unperturbed cavity eigenfrequency?
We answer this question in this brief report.

If the selected field mode interacts with the $N$-level detector in
resonant ladder configuration,
the Hamiltonian describing the whole system ``field mode + atom''
can be taken in the Rotating Wave Approximation (RWA) form
\begin{equation}
H=H_{c}+\sum_{i=1}^{N}E_{i}\sigma _{ii}+\sum_{i=1}^{N-1}g_{i}(a\sigma
_{i+1,i}+a^{\dagger }\sigma _{i,i+1}),
\label{totHam}
\end{equation}%
where $E_{i}$ is the energy of the $i$-th atomic eigenstate
$|\mathbf{i}\rangle $ (in bold), $\sigma _{i,j}\equiv |\mathbf{%
i}\rangle \langle \mathbf{j}|$ is the generalized Pauli operator,
and $g_{i}$ (assumed to be real) is the
coupling parameter between the atomic states \{$|\mathbf{i}\rangle ,|\mathbf{%
i}+\mathbf{1}\rangle $\} through the cavity field.
The condition of validity of RWA is $|g_j|\ll 1$. We consider here the resonant case,
so we assume $E_{i+1}-E_{i}=1$ for $i=1,\dots ,N-1$. Our aim is
to find out how the photon generation \emph{from vacuum} is affected by the
detector in the weak modulation regime, $|\varepsilon|\ll g_{j}$ ($j=1,\dots ,N-1$)
for the modulation frequency $\eta =2$.

To find the wave function of the whole system $|\Psi(t)\rangle$,
we make the transformation $|\Psi (t)\rangle
=V(t)|\psi (t)\rangle $  with $V(t)=\exp
[-it(n+\sum_{i=1}^{N}E_{i}\sigma _{ii})]$. Then after RWA we get the
following Hamiltonian governing the time evolution of $|\psi (t)\rangle $
in the interaction picture (where $\beta\equiv {\varepsilon }/{4}$):
\begin{equation}
H_{I}=-i\beta\left(a^{2}-a^{\dagger 2}\right)
+\sum_{i=1}^{N-1}g_{i}(a\sigma _{i+1,i}+a^\dagger \sigma _{i,i+1}).
 \label{iron}
\end{equation}

In the special case when $N\rightarrow \infty $ and $g_{j}=%
\sqrt{j}g$ the detector becomes a simple harmonic oscillator (HO) if we
associate the atomic level $|\mathbf{1}\rangle $
with the oscillator ground  (zero energy) state  and make the replacements
\[
\sum_{j=1}^{\infty} \sqrt{j}\sigma _{j+1,j}=b^{\dagger },
\quad \sum_{j=1}^{\infty} \sqrt{j}\sigma _{j,j+1}=b ,
\]
 where $b$
and $b^{\dagger }$ are the annihilation and creation operators associated with
the detector ($[b,b^{\dagger }]=1$). Then the Heisenberg equations of motion
for operators $a(t)$ and $b(t)$ can be solved exactly:
\begin{eqnarray}
a(t) &=& \left(C_{\beta}c_{\gamma} + \beta S_{\beta}s_{\gamma}\right) a_0
-ig C_{\beta}s_{\gamma} b_0
\nonumber \\ &+&
\left(S_{\beta}c_{\gamma} + \beta C_{\beta}s_{\gamma}\right) a_0^{\dagger}
+ig S_{\beta}s_{\gamma} b_0^{\dagger},
\label{at}
\end{eqnarray}
\begin{eqnarray}
b(t) &=& \left(C_{\beta}c_{\gamma} - \beta S_{\beta}s_{\gamma}\right) b_0
-ig C_{\beta}s_{\gamma} a_0
\nonumber \\ &-&
\left(S_{\beta}c_{\gamma} - \beta C_{\beta}s_{\gamma}\right) b_0^{\dagger}
-ig S_{\beta}s_{\gamma} a_0^{\dagger},
\label{bt}
\end{eqnarray}
where $\gamma =\sqrt{g^2 -\beta^2}$, $C_{\beta} \equiv \cosh(\beta t)$,
$ S_{\beta} \equiv \sinh(\beta t)$, $ c_{\gamma} \equiv \cos(\gamma t)$,
and $ s_{\gamma} \equiv \sin(\gamma t)/\gamma$.
If $|\beta|>|g|$, then trigonometrical functions should be replaced by
their hyperbolic counterparts with $\gamma$ replaced by
$\tilde\gamma =\sqrt{\beta^2 - g^2 }$.
The mean number of quanta in the field mode $\langle n(t)\rangle$
and the mean excitation number of the detector $\langle n_b(t)\rangle$
 for the initial vacuum state are equal to
\begin{equation}
\left.
\begin{array}{c}
\langle n(t)\rangle \\
\langle n_b(t)\rangle
\end{array}
\right\}
=
S_{\beta}^2 \pm\beta S_{2\beta} s_{2\gamma} +\beta^2 C_{2\beta} s_{\gamma}^2\,.
\label{nt}
\end{equation}
The Mandel factor and quadrature variances of the field mode are
\begin{equation}
Q(t) = \langle n(t)\rangle
+\frac{\left[S_{2\beta}\left(1+2\beta^2 s_{\gamma}^2\right)
+ 2\beta C_{2\beta} s_{2\gamma}\right]^2}{4\langle n(t)\rangle}\,,
\label{Qt}
\end{equation}
\begin{equation}
\left.
\begin{array}{c}
(\Delta p)^{2} \\
(\Delta x)^{2}
\end{array}
\right\}
=
 e^{\mp2\beta t}\left(\frac{1}{2} \mp\beta s_{2\gamma} +\beta^2 s_{\gamma}^2\right).
  \label{pt}
\end{equation}
If $g=0$, then Eqs. (\ref{nt})-(\ref{pt}) go to (\ref{n0})-(\ref{p0}).
But if $|g|\gg |\beta|$, then the rate of photon generation becomes roughly twice smaller
than in the empty cavity
\footnote{%
A similar effect was discovered for empty cavities with additional symmetry
(such as cubical ones), when several cavity modes can be in resonance with
the external perturbations and between themselves \cite{Croc1,AVD}: in this
case other modes play the role of an effective oscillator ``detector''.}.
Moreover, the ratio $\langle n_b(t)\rangle/\langle n(t)\rangle$ is close
to unity if $|g| \gg |\beta|$.
For $\varepsilon>0$ and $\beta t \gg 1$ we see the exponential growth
of the mean photon number with increment $\varepsilon/2$, which is modulated by some oscillations
with the frequency $2\gamma$:
\begin{equation}
\langle n(t)\rangle  \approx \frac14 e^{2\beta t}\left[ 1
+ \frac{\beta}{\gamma} \sin(2\gamma t) + \frac{2\beta^2}{\gamma^2}\sin^2(\gamma t)\right].
\label{steps}
\end{equation}
For $t=t_n +\delta t$ with $2\gamma t_n=(2n+1)\pi$ we have the following Taylor expansion of
formula (\ref{steps}):
\[
\langle n(t_n +\delta t)\rangle  \approx \langle n(t_n)\rangle
\left[1 +(4/3)\beta g^2(\delta t)^3\right].
\]
Consequently, the function $\langle n(t)\rangle $ is practically constant in some neighborhood
of $t_n$. This results in the appearance of almost horizontal ``shelves'' in the plots of
$\langle n(t)\rangle $, $Q(t)$ and $(\Delta p)^{2}(t)$,
which are clearly seen in Figs. \ref{f1} and \ref{f2} below.
For $\beta t \gg 1$ Eq. (\ref{Qt}) can be simplified as $Q(t)\approx 2\langle n(t)\rangle$,
which is a typical relation for highly squeezed vacuum states. Nonetheless, the state of the
field mode is not exactly the vacuum squeezed one, since the uncertainty product
$\Delta \equiv (\Delta p)^{2}(\Delta x)^{2}$ is bigger than the minimal possible value $1/4$:
$\Delta=1/4 + g^2 \beta^2 s_{\gamma}^4$ (the covariance between $x$ and $p$ quadratures is zero
in the case discussed). As soon as the initial state of the system was Gaussian,
it remains Gaussian for all times. In such a case, the purity of the field mode equals
\cite{D-book} $\mu \equiv \mbox{Tr}(\hat\rho^2)= (4\Delta)^{-1/2}$,
where $\hat\rho$ is the statistical operator. If $|g|\gg |\beta|$,
the purity is only slightly below unity, oscillating with the frequency $2\gamma$.
But for $0< |g| < |\beta|$ the purity goes monotonously to zero.
\begin{figure}[tbh]
\begin{center}
\includegraphics[width=0.49\textwidth]{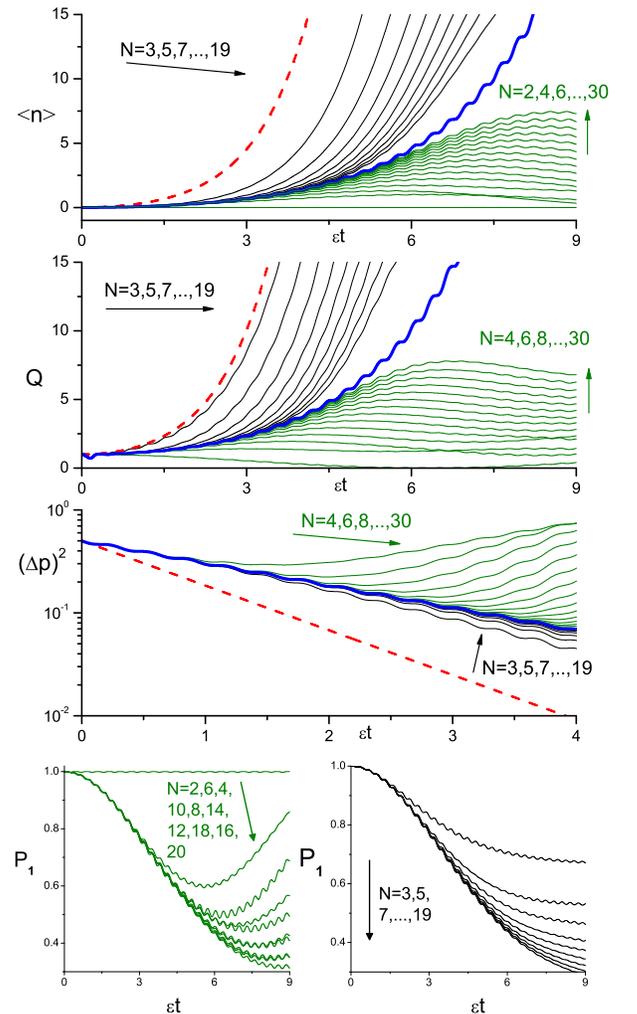} {}
\end{center}
\par
\vspace{-1cm}
\caption{(Color online) Various quantities as functions of the dimensionless
time $\varepsilon t$ for different
numbers of the detector levels for $g_{j}=\protect\sqrt{j}g$ with
$g=10^{-2}$ and $\varepsilon =10^{-3}$. Blue thick lines
represent the results for harmonic oscillator detector, while the red dashed
line --  for the empty cavity.}
\label{f1}
\end{figure}

For equidistant detectors with a finite number of levels analytical solutions
cannot be obtained. In this case
we solve the Schr\"{o}dinger equation for the Hamiltonian (\ref{iron}) by
expanding the wavefunction in the Fock basis as $|\psi (t)\rangle
=\sum_{j=1}^{N}\sum_{k=0}^{\infty }p_{j,k}(t)|\mathbf{j},k\rangle $, where
the first index stands for the atomic eigenstate and the second for the Fock
state of the cavity field. We get the set of differential
equations%
\begin{eqnarray}
\dot{p}_{j,m} &\!\!=\!\!& \beta(\sqrt{m(m-1)}p_{j,m-2}-\sqrt{(m+1)(m+2)}p_{j,m+2})  \notag
\\&&
-i(g_{j}\sqrt{m}p_{j+1,m-1}+g_{j-1}\sqrt{m+1}p_{j-1,m+1})
\label{megadeth}
\end{eqnarray}%
where $g_{0}=g_{j\geq N}\equiv 0$
and the dot stands for the time derivative.
Hereafter we suppose that $|\beta|\ll |g_j|$.
To
understand qualitatively the resulting dynamics in this case we follow the method
employed in \cite{1,4,5}. First we solve the equations (\ref{megadeth}) for $%
\beta=0$, obtaining solutions in the form of
exponentials with time-varying arguments multiplied by constant coefficients. Then
we substitute the  solutions obtained back into (\ref{megadeth}) with $\beta\neq
0$, assuming that now the coefficients at the exponentials become
time-dependent. Thus we obtain a set of differential equations for
these coefficients.

In this paper we assume that the only nonzero initial probability amplitude
 is $p_{1,0}(0)=1$, so the first equation to be solved is $\dot{%
p}_{1,0}(t)=-\sqrt{2}\beta p_{1,2}(t)$. Our first task is to find the expression
for $p_{1,2}(t)$ when $\beta=0$; it must be obtained from the matrix equation%
\begin{equation}
\mathbf{\dot{X}}=-i\mathbf{MX~},  \label{metallica}
\end{equation}%
where $\mathbf{X}\equiv (p_{1,2};p_{2,1};p_{3,0})$ and
\begin{equation}
\mathbf{M}=\left(
\begin{array}{ccc}
0 & \sqrt{2}g_{1} & 0 \\
\sqrt{2}g_{1} & 0 & g_{2} \\
0 & g_{2} & 0%
\end{array}%
\right) .  \label{motorhead}
\end{equation}%
For $N=2$ the eigenvalues of matrix $\mathbf{M}$ are $\pm \sqrt{2}g_{1}$, so
  $p_{1,2}=\sum_{k=1}^{2}\Lambda_{k}^{(2)}\exp (-i\varphi _{k}^{(2)}t)$,
where $\Lambda _{k}^{(2)}$ are constant coefficients and
$\varphi^{(2)} _{k}$ are the
eigenvalues of $\mathbf{M}$. Since these eigenvalues are big, $|\varphi^{(2)}
_{j}|\gg |\beta|$, no significant coupling between the coefficients
$p_{1,0}$ and $\Lambda _{k}^{(2)}$ (with $k=1,2$) arises
when one puts the ansatz $p_{1,2}=\sum_{k=1}^{2}\Lambda
_{k}^{(2)}(t)\exp (-i\varphi _{k}^{(2)}t)$ into Eq. (\ref{megadeth}) with $%
\beta\neq 0$. So within the RWA reasoning one has $%
\dot{p}_{1,0}\simeq 0$. Thus $p_{1,0}$ is decoupled from all the other
probability amplitudes and the system remains in the initial state $|\mathbf{1%
},0\rangle $. On the other hand, for $N>2$ the solution of (\ref{metallica})
for $p_{1,2}$ is $p_{1,2}=\sum_{k=1}^{3}\Lambda _{k}^{(3)}\exp (-i\varphi
_{k}^{(3)}t)$ with the third eigenvalue of matrix $\mathbf{M}$ equal to
zero, $\varphi _{3}^{(3)}=0$. Moreover, one can check that $p_{2,1}$ does not
contain the term $\Lambda _{3}^{(3)}$, but $p_{3,0}$ does (exact
expressions are given in \cite{4}). Hence, in this case $p_{1,0}$ is
resonantly coupled to $\Lambda _{3}^{(3)}$, so $p_{1,2}$ and $p_{3,0}$
become populated, while $p_{2,1}$ is coupled off-resonantly, so it is much
smaller than the former two coefficients ($|\Lambda _{1}^{(3)}|,|\Lambda _{2}^{(3)}|\ll
|\Lambda _{3}^{(3)}|$). Thus in this case at least two photons can be created for sure.

Now we must see the coupling of the coefficients $\Lambda _{j}^{(3)}$ ($%
j=1,2,3$) to the next subset, so we have to solve Eq. (\ref%
{metallica}) with $\mathbf{X}=(p_{1,4};p_{2,3};p_{3,2};p_{4,1};p_{5,0})$ and
\begin{equation*}
\mathbf{M}=\left(
\begin{array}{ccccc}
0 & g_{1}\sqrt{4} & 0 & 0 & 0 \\
g_{1}\sqrt{4} & 0 & g_{2}\sqrt{3} & 0 & 0 \\
0 & g_{2}\sqrt{3} & 0 & g_{3}\sqrt{2} & 0 \\
0 & 0 & g_{3}\sqrt{2} & 0 & g_{4} \\
0 & 0 & 0 & g_{4} & 0%
\end{array}%
\right)
\end{equation*}%
For $N=4$ we have $p_{1,4}=\sum_{k=1}^{4}\Lambda
_{k}^{(4)}\exp (-i\varphi _{k}^{(4)}t)$, where all the eigenvalues are
different from zero and much larger than $|\beta|$,
being functions of $g_{i}$ ($i=1,2,3$). Therefore $\Lambda _{3}^{(3)}$ does not couple
resonantly to any $\Lambda _{k}^{(4)}$. Thus at most two photons are
generated for the four-level atom. For $N>4$ the solution for $p_{1,4}$ is $%
p_{1,4}=\sum_{k=1}^{5}\Lambda _{k}^{(5)}\exp (-i\varphi _{k}^{(5)}t)$ with
$\varphi _{5}^{(5)}=0$, and $\Lambda _{5}^{(5)}$ also appears in $p_{3,2}$
and $p_{5,0}$, but not in $p_{2,3}$ and $p_{4,1}$. Thus, the coefficient $%
\Lambda _{3}^{(3)}$ is resonantly coupled to $\Lambda _{5}^{(5)}$ and at
least four photons can be generated.

Continuing this analysis for higher order matrices, one can easily verify
that matrices $\mathbf{M}$ of odd orders always have one null
eigenvalue, and the coefficient $\Lambda $ that multiplies the exponential
with the null eigenvalue appears in probability amplitudes with odd atomic
level and even photon number. On the other hand,
all eigenvalues of matrices $\mathbf{M}$ of even orders  are different from
zero and much larger than $|\beta|$.
Hence we arrive at the general rule:
\emph{The number of created photons is unlimited for odd numbers of levels,
while at most $N-2$ photons can be created if the number of levels $N$ is even.}
 Moreover, the
probability of detecting an odd number of photons is much smaller than the
probability of detecting an even number of photons.

These results can be  extended straightforwardly to another practical
scenario:  %for detecting Casimir photons in cavities:
$\left( N-1\right) $
resonant 2-level atoms interacting with the field mode, when initially all
the atoms are in their ground states and the field is in the vacuum state,
for the modulation frequency $\eta =2$. It was shown in \cite{5} that for a
single atom no photons are generated for this modulation frequency, and for
two atoms the photons are generated steadily provided the coupling strengths
are equal. So here we assume that all the couplings are equal to $g$. In the
interaction picture the Hamiltonian describing this setup is the
Tavis-Cummings Hamiltonian \cite{Tavis,5} with additional parametric
amplification term analogous to Eq. (\ref{iron})
\begin{equation}
H_{I}=-i\beta(a^{2}-a^{\dagger 2})+g(aS_{+}+a^{\dagger
}S_{-}),  \label{lemmy}
\end{equation}%
where $S_{\pm }=\sum_{j=1}^{N-1}\sigma _{\pm }^{(j)}$ are the collective
ladder operators and $\sigma _{j}^{-}=|\mathbf{1}_{j}\rangle \langle \mathbf{%
2}_{j}|$, $\sigma _{j}^{+}=|\mathbf{2}_{j}\rangle \langle \mathbf{1}_{j}|$
are the standard 2-level Pauli operators, where $|\mathbf{1}_{j}\rangle $
and $|\mathbf{2}_{j}\rangle $ are the ground and excited states of the $j$%
-th atom ($j=1,\dots ,N-1$), respectively. Writing the wavefunction
associated to the Hamiltonian (\ref{lemmy}) as $|\psi (t)\rangle
=\sum_{j=1}^{N}\sum_{k=0}^{\infty }p_{j,k}|\mathbf{j},k\rangle $, where $|%
\mathbf{j}\rangle $ denotes the symmetric normalized Dicke state \cite{Dicke}
with $\left( j-1\right) $ excitations ($j=1,\dots ,N$) \footnote{%
Notice that we can use the Dicke states \emph{only} when all
coupling coefficients are equal.} and $|k\rangle $ is the cavity Fock state,
and using the known properties $S_{+}|\mathbf{j}\rangle =\sqrt{j\left(
N-j\right) }|\mathbf{j+1}\rangle $ and $S_{-}|\mathbf{j}\rangle =\sqrt{%
\left( j-1\right) \left( N-j+1\right) }|\mathbf{j-1}\rangle $, we obtain
precisely the equation (\ref{megadeth}) for the probability amplitudes,
where the effective coupling strengths are $g_{j}\equiv g\sqrt{j\left(
N-j\right) }$ ($j=1,\dots ,N$). Thus our statement holds unaltered.
\begin{figure}[tbh]
\begin{center}
\includegraphics[width=0.49\textwidth]{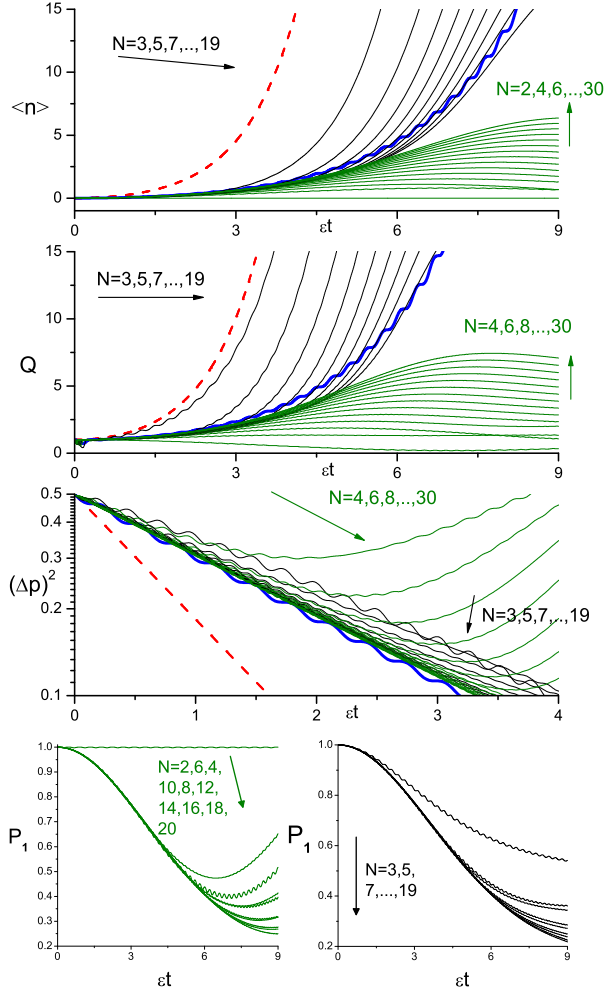} {}
\end{center}
\par
\vspace{-1cm}
\caption{(Color online) Same as Fig. \protect\ref{f1} for $g_{j}=\protect%
\sqrt{j\left( N-j\right) }g$.}
\label{f2}
\end{figure}
\begin{figure}[tbh]
\begin{center}
\includegraphics[width=0.49\textwidth]{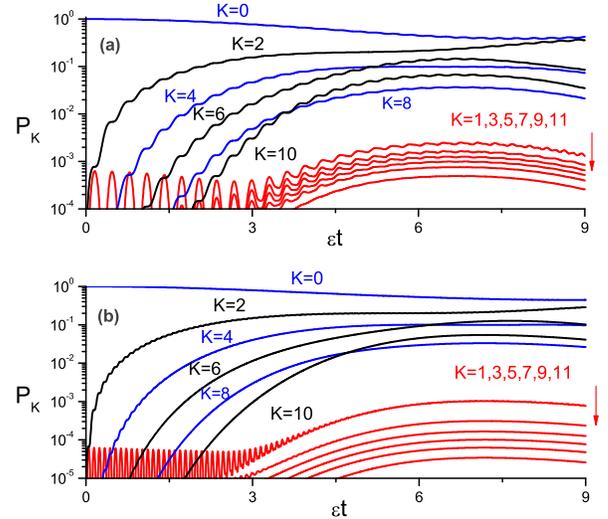} {}
\end{center}
\par
\vspace{-1cm}
\caption{(Color online) Probability $P_{K}$ of detecting $K$ photons as
function of dimensionless time $\varepsilon t$ for $N=12$,
$g=10^{-2}$ and $\varepsilon =10^{-3}$.
\textbf{a)} $g_{j}=\protect\sqrt{j}g$. \textbf{%
b)} $g_{j}=\protect\sqrt{j\left( N-j\right) }g$.
}
\label{f5}
\end{figure}

We confirmed these results by performing numerical simulations in which we
solved the equations (\ref{megadeth}) using the Runge-Kutta-Verner
fifth-order and sixth-order method for different forms of the coupling
strengths
\footnote{The precision of numerical calculations was verified
by calculating the total sum of probabilities.
In all the cases it was equal to unity with errors below the level of $10^{-10}$.
}. In Fig. \ref{f1} we show the behavior of the mean photon
number $\langle n\rangle $, the Mandel factor $Q$, the squeezed quadrature
variance $(\Delta p)^{2}$ and the probability $P_{\mathbf{1}}$ of
finding the detector in the ground state as function of dimensionless time $%
\varepsilon t$ for different number of detector's levels $N$. We verified
that the probability of finding the detector in an even energy level is
always very low (data not shown). In Fig. \ref{f1} we set $g_{j}=\sqrt{%
j}g$ ($j=1,\dots ,N-1$), as one expects that for $N\gg 1$ the results should
approach the behavior of the case of harmonic oscillator detector (depicted
with thick blue line), and the red dashed lines denote
the results for the empty cavity according to Eqs. (\ref{n0})--(\ref{p0}).
In Fig. \ref{f2} we set $g_{j}=\sqrt{j\left(
N-j\right) }g$ corresponding to $N-1$ two-level atoms.
%The numerical values are $g=10^{-2}$ and $\varepsilon =10^{-3}$,
In Fig. \ref{f5} we plot the probabilities $P_{K}$ of detecting $K$
photons as function of time for $g_{j}=\sqrt{j}g$ (Fig. \ref{f5}a) and $%
g_{j}=\sqrt{j\left( N-j\right) }g$ (Fig. \ref{f5}b) for $N=12$, to show that
at most $10$ photons are created, whereas the probabilities of detecting odd
photon numbers are very small (the probabilities with $K\ge 12$ are below the horizontal axes).
Hence the numerical results are in full agreement with our general rule.

In conclusion, we studied the photon pair creation from vacuum via the
Dynamical Casimir Effect in a cavity containing a resonant $N$-level
detector in equally-spaced ladder configuration or $N-1$ identical resonant
two-level atoms. We considered the regime of weak modulation, where the
cavity modulation depth is much smaller than the atom-field coupling. We
showed that for an odd number of levels pairs of photons are generated
steadily as time goes on, while for an even number of levels at most $N-2$
photons are created. For $N\to \infty$ the difference between even and
odd values of $N$ eventually disappears, and for the special choice of the
coupling constants we arrived at the harmonic oscillator model
of the detector. In this special case the rate of photon generation is roughly
twice smaller than for an empty cavity.
 These findings can be useful for constructing efficient
detectors for monitoring the photon generation via Dynamical Casimir Effect,
because the detector can be excited without interrupting the photon growth.

\begin{acknowledgments}
V.V.D. acknowledges the partial support of CNPq (Brazilian agency).
\end{acknowledgments}


\begin{thebibliography}{99}

\bibitem{revDCE} V. V. Dodonov, Phys. Scr. \textbf{82}, 038105 (2010).

\bibitem{revDal} D. A. R. Dalvit, P. A. Maia Neto, and F. D. Mazzitelli,
%Fluctuations, dissipation and the dynamical Casimir effect,
in \emph{Casimir Physics}, edited by D. Dalvit, P. Milonni, D. Roberts, and
F. da Rosa, Lecture Notes in Physics Vol. 834 (Springer, Berlin, 2011), p.
419. %[arXiv: 1006.4790].

\bibitem{revRMP} P. D. Nation, J. R. Johansson, M. P. Blencowe, and F. Nori,
Rev. Mod. Phys. \textbf{84}, 1 (2012).

\bibitem{DK96} V. V. Dodonov and A. B. Klimov, Phys. Rev. A \textbf{53},
2664 (1996); %
%\bibitem{Plunien}
G. Plunien, R. Sch\"utzhold, and G. Soff, Phys. Rev. Lett. \textbf{84}, 1882
(2000); %
%\bibitem{Croc1}
M. Crocce, D. A. R. Dalvit, and F. D. Mazzitelli,
%Resonant photon creation in a three-dimensional oscillating cavity
Phys. Rev. A \textbf{64}, 013808 (2001).

\bibitem{Law94} C. K. Law, Phys. Rev. A \textbf{49}, 433 (1994).


\bibitem{pla} V. V. Dodonov, Phys. Lett. A \textbf{207}, 126 (1995).

\bibitem{1} A. V. Dodonov, R. Lo Nardo, R. Migliore, A. Messina, and V. V.
Dodonov, J. Phys. B \textbf{44}, 225502 (2011).

\bibitem{2} A. V. Dodonov and V. V. Dodonov, Phys. Lett. A \textbf{375},
4261 (2011);
%
%\bibitem{3} A. V. Dodonov and V. V. Dodonov,
Phys. Rev. A \textbf{85}, 015805 (2012).

\bibitem{4} A. V. Dodonov and V. V. Dodonov, arXiv:1202.0772.

\bibitem{5} A. V. Dodonov and V. V. Dodonov, arXiv:1203.3776.

\bibitem{Croc1} M. Crocce, D. A. R. Dalvit, and F. D. Mazzitelli,
%Resonant photon creation in a three-dimensional oscillating cavity
Phys. Rev. A {\bf 64},  013808 (2001).

\bibitem{AVD} A. V. Dodonov and V. V. Dodonov, Phys. Lett. A {\bf 289},
291 (2001).
%Nonstationary Casimir effect in cavities with two resonantly coupled modes


\bibitem{D-book} V. V. Dodonov, in \emph{Theory of Non-classical States of Light},
edited by V. V. Dodonov and V. I. Man'ko (Taylor \& Francis, London, 2003), p. 153.


\bibitem{Tavis} M. Tavis and F. W. Cummings, Phys. Rev. \textbf{170}, 379
(1968).

\bibitem{Dicke} R. Dicke, Phys. Rev. \textbf{93,} 99 (1954).
\end{thebibliography}
\end{document}